\documentstyle[aps,prl]{revtex}
\input{epsf}

\begin{document}
\title{An efficient joint source-channel coding for a D-dimensional
array} \author{Ido Kanter, Haggai Kfir and  Shahar Keren} 
\address{Minerva Center and the
Department of Physics, Bar-Ilan University, Ramat-Gan 52900, Israel 
}

\date{July 2003}
\maketitle

\begin{abstract}
An efficient joint source-channel (S/C) decoder based on the side
information of the source and on the MN-Gallager Code over Galois
fields, $q$, is presented. The dynamical posterior probabilities are
derived either from the statistical mechanical approach for
calculation of the entropy for the correlated sequences, or by the
Markovian joint S/C algorithm.  The Markovian joint S/C decoder has
many advantages over the statistical mechanical approach, among them:
(a) there is no need for the construction and the diagonalization of a
$q \times q$ matrix and for a solution to saddle point equations in
$q$ dimensions; (b) a generalization to a joint S/C coding of an array
of two-dimensional bits (or higher dimensions) is achievable; (c)
using parametric estimation, an efficient joint S/C decoder with the
lack of side information is discussed. Besides the variant joint S/C
decoders presented, we also show that the available sets of
autocorrelations consist of a convex volume, and its structure can be
found using the Simplex algorithm.
\end{abstract}


%

\section{Introduction}
Source coding is a process for removing redundant information from the
source information symbol stream. Suppose we have a bitmapped image,
then converting the bitmap image to GIF, JPEG or any of the familiar
image formats used on the web is a source coding process. Not only
can images be coded, but also sound, video frames, etc., and
compressing the stream of information is source coding.

Channel coding is a procedure for adding redundancy as protection into
the information stream which is to be transmitted; in other words,
channel coding can be regarded as adding protection to the
transmission process.  For example, a wireless communication channel
is affected by many factors such as distance, speed at
which either party is moving, weather, buildings, other users'
unintentional interference, etc., so errors cannot be avoided.  During
the last decade engineers and also physicists have designed efficient
error correction techniques such as Low-Density-Parity-Check-Codes
(LDPC)\cite{forney,David-Mackay2,Shokrollahi,KS} or Turbo\cite{turbo} codes
that nearly saturate Shannon's limit.

In a typical scenario of a communication channel there are two major
resources which are highly limited. The first is power, which
includes both transmitter power and receiver power.  The
second is bandwidth (channel capacity) indicating the speed at which 
the channel can transmit information, or more exactly, how many bps (bits
per second). Both of these determine the capability of a
channel. For example, by increasing the power we can reduce the error,
but the power is limited. On the other hand, if the channel capacity
is unlimited, we can just go ahead and add a large amount of protection
(low rate), but again we cannot afford that since channel capacity is a
commodity which in many scenarios is even more precious than power.

The main tradeoff in communication is the following: given a
fixed capacity channel and a fixed amount of power, how should we
allocate them between the source and the channel to get the best
result, i.e, the smallest distortion?  We know that a certain amount of
channel capacity is allocated to the source and the rest is used for
protection, but what is the ratio between them?

Shannon separation theorem states that source coding (compression)
and channel coding (error protection) can be performed separately and
sequentially, while maintaining optimality
\cite{Shannon,Cover,err_cor_book,Frey}. However, this is true only in
the case of asymptotically long block lengths of data and
point-to-point transmission. In many practical applications, the
conditions of the Shannon's separation theorem neither holds, nor can
it be used as a good approximation. Thus, considerable interest has
developed in various schemes of joint source-channel (S/C) coding,
where compression and error correction are combined into one mechanism
(see, for instance, the following selected publications
\cite{shamail1,shamail2,shamail3,shamail4,shamail5,shamail6}).



The paper is organized as follows.
In Section II Statistical Mechanical (SM) joint S/C coding is
introduced, whereas in Section III the threshold of the code is
calculated using scaling behavior for the required number of messages
passing for the convergence of the algorithm
\cite{KS,KS-Gaussian,KK}. In Section IV the efficiency of the SM
joint S/C coding is compared to various separation schemes. A
degradation in the performance of the SM joint S/C coding is examined
in Section V as a function of the spectrum of the eigenvalues of the
transfer matrix. In Section VI the Simplex algorithm is used to
calculate the available space of a possible set of autocorrelations.
The drawbacks of the SM joint S/C coding are discussed in Section VII,
and advanced S/C coding is presented in Section VIII. The Markovian
joint S/C coding and its efficiency are discussed in Section IX. Based
on the parametric estimation methods the Markovian joint S/C decoder
with the lack of side information is discussed in Section X.  Its
extension to higher dimensions is discussed in Section XI. The paper 
closes with some concluding remarks.

\section{Joint S/C coding - Statistical Mechanical approach}

In our recent papers\cite{KR,KK} a particular scheme based on a
SM approach for the implementation of the joint
S/C coding was presented and the main steps are briefly
summarized below.  The original boolean source is first mapped to a
binary source \cite{sourlas,sourlas1} $\left\{ x_{i}\pm1\right\}
~i=1,...,L$, and is characterized by a finite set of autocorrelations
bounded by the length $k_0$
\begin{equation}
C_{k_1, ...,k_m}=\frac{1}{L}\sum_{i=1}^{L}x_{i}\prod_{j=0}^m
x_{\left(i+k_j\right)\: \mathbf{mod}\: L}
\label{ck}
\end{equation}
\noindent where $k_m \le k_0$ is the highest length autocorrelation
taken and the total number of possible different autocorrelations is
$2^{k_0}$.  For $k_0=2$, for instance, there are only $4$ possible
correlations, $C_0$, $C_1$, $C_2$ and $C_{12}$, and for $k_0=3$ there
are $8$ possible different correlations;
$C_0,~C_1~,C_2,~C_3,~C_{12},~C_{13},~C_{23},~C_{123}$, where we do not
assume left-right symmetry for the source.  Note that for the general
$k_0$ and $m=1$, eq. \ref{ck} is reduced to the two-point
autocorrelation functions \cite{liat}. The number of sequences obeying
these $2^{k_0}$ constraints is given by
\begin{equation}
\Omega = Tr_{\{ x_i = \pm 1 \} }\!\!\!\! 
\prod_{\{k_1,k_2,...,k_m\}}\!\!\!\!\! \delta
(\sum_{i=1}^{L}x_{i}\prod_{j=0}^m x_{i+k_j} - LC_{k_1,
...,k_m})
\label{omega}
\end{equation}
\noindent where $m=0$ stands for $C_0$. Using the integral
representation of the delta functions, eq. \ref{omega} can be written
as
\begin{eqnarray}
&\Omega\!\! &\!\! =\!\! \int_{-\infty}^{\infty}
\prod_{\{k_1,..,k_m\}}\!\!\!\!
dy_{\{k_1,..,k_m\}} \exp(\sum
-y_{k_1,..,k_m}C_{k_1,..,k_m}) \nonumber  \\ & Tr &\!\!\!\! \exp (\sum_{k_1,..,k_m}
y_{k_1,...,k_m}\sum_i x_{i}\prod_{j=0}^m x_{i+k_j} )
\label{omega1}
\end{eqnarray}
Since $k_j \le k_0$, the last term of eq. \ref{omega1} indicates that
the trace can be performed using the standard transfer matrix (of size
$2^{k_0} \times 2^{k_0}$) method\cite{baxter}. More precisely, assume
two successive blocks of $k_0$ binary variables denoted by
$(x_1,...,x_{k_0})$ and $(x_{k_0+1},...  ,x_{2k_0})$. The element
$(i,j)$ of the transfer matrix is equal to the value of the last
exponential term (on the r.h.s of the trace) of eq. \ref{omega1},
where the first block is in state $i$ (among $2^{k_0}$ possible
states) and the second block is in state $j$. The transfer matrix is a
non-negative matrix (as long as the $y_{k_1,...,k_m}$ are real
numbers), and the leading eigenvalue is positive and
non-degenerate\cite{baxter}. In the leading order one finds
\begin{eqnarray}
\Omega &=&\int
dy_k\exp\{-L\lbrack \sum y_{k_1,...,k_m}C_{k_1,...,k_m} \nonumber \\
&-& \ln \lambda_{max}(\{ y_{k_1,...,k_m}\})
\rbrack \}
\label{omega-sp}
\end{eqnarray} 
\noindent where $\lambda_{max}$ is the maximal eigenvalue of the
corresponding transfer matrix.  For large $L$ and using the saddle
point method, the entropy, $H_2(\{C_{k_1,...,k_m} \})$, is given in the
leading order by
\begin{eqnarray}
H_2\left(\{C_{k_1...,k_m}\}\right) &=& {1 \over \ln 2} \lbrack 
\frac{1}{k_0}\ln
\lambda_{max}\left (\{
y_{k_1,...,k_m}\}\right) \nonumber \\
&-&\sum_{k_1,...,k_m}^{k_0}y_{k_1,...,k_m}
C_{k_1,...,k_m} \rbrack
\label{entropy-ck}
\label{h2}
\end{eqnarray}
\noindent where $\{y_{k_1,...,k_m}\}$ are determined from the saddle
point equations of $\Omega$\cite{KK}. Assuming Binary Symmetric
Channel (BSC) and using Shannon's lower bound, the channel capacity of
sequences with a given set of autocorrelations bounded by a distance
$k_0$ is given by
\begin{equation}
C=\frac{1-H_{2}\left(f\right)}{H_{2}(\left\{C_{k_1,...,k_m}\}\right)-
H_{2}\left(P_{b}\right)}
\label{capacity}
\end{equation}
\noindent where $f$ is the channel bit error rate and $p_b$ is a bit
error rate.  The saddle point solutions derived from
eq. \ref{omega-sp} indicate that the equilibrium properties of the
one-dimensional Ising spin system ($x_i=\pm1$) with up to order $k_0$
multi-spin interactions\cite{ido-msi}
\begin{equation}
H=-\sum_i \sum_{k=1}^{k_0} \frac{y_{k_1,...,k_m}}{\beta }
x_{i}\prod_{j=0}^m x_{i+k_j} 
\label{hamiltonian}
\end{equation}
\noindent obey in the leading order the autocorrelation constraints,
eq. \ref{ck}.  This property of the effective Hamiltonian,
eq. \ref{hamiltonian}, is used in simulations to generate an ensemble of
signals (source messages) with the desired set of
autocorrelations. {\it Note that in the following we choose $\beta=1$,
and hence we denote $\{y_{k_1,...,k_m}\}$ as interactions.}


The transfer matrix method indicates that the relevant scale of the
correlated source message is $k_0$.  Hence, our encoding/decoding
procedure is based on the MN code\cite{MacKay} for a finite field
$q=2^{k_0}$ \cite{LDPC-GF(q),Davey}, which is based on the
construction of two sparse matrices $A$ and $B$ of dimensionalities
$L_0/R\!\times\! L_0$ and $L_0/R\! \times\! L_0/R$ respectively, where
$R$ is the code-rate and $L_0=L/k_0$.
The matrix $B^{-1}A$ is then used for encoding the message
\begin{equation}
t = B^{-1}A x\  (\: \mathbf{mod}\:\ \ q)
\label{trans}
\end{equation}
The finite field message vector $t$ is mapped to a binary vector and
then transmitted. The received message, $r$, is corrupted by the
channel bit error rate, $f$.

The decoding of symbols of $k_0$ successive bits (named in the
following as a {\it block} of bits or binary variables) is based on the
solution of the syndrome
\begin{equation} 
Z= Br = Ax + Bn\   (\: \mathbf{mod}\:\ \ q)
\label{decoding}
\end{equation}
\noindent where $n$ stands for the corresponding noise of $k_0$
successive bits.  The solution of the $L_0/R$ equations with
$L_0(1/R+1)$ variables is based on the standard message passing
introduced for the MN decoder over Galois fields with
$q=2^{k_0}$\cite{LDPC-GF(q),Davey} and with the following modification. The
horizontal pass is left unchanged, {\it but a dynamical set of
probabilities assigned for each block is used in the vertical
pass}. The Dynamical Block Probabilities (DBP), $\{P_n^c\}$, are
determined following the current belief regarding the neighboring
blocks and are given by
\begin{eqnarray}
\gamma_{n}^{c} & = & S_{I}\left(c\right)\left(\sum
_{l=1}^{q}q_{L}^{l}S_{L}\left(l,c\right)\right)\left(\sum
_{r=1}^{q}q_{R}^{r}S_{R}\left(c,r\right)\right)\nonumber \\ P_{n}^{c}
& = & \frac{\gamma _{n}^{c}}{\sum _{j=1}^{q}\gamma
_{n}^{j}}\label{tm-vertical-pass}
\label{dbp}
\end{eqnarray}
\noindent where $l/r/c$ denotes the state of the left/right/center
($n\!-\!1\,/\,n\!+\!1\,/\,n$) block respectively and
$q_{L}^{l}/q_{R}^{r}$ are their posterior probabilities.
$S_I(c)=e^{-\beta H_I}$ stands for the Gibbs factor of the inner
energy of a block, $k_0$ successive binary variables spins,
characterized by an energy $H_I$ at a state $c$, see
eq. \ref{hamiltonian}.
Similarly $S_L(l,c)$ ($S_R(c,r)$) stands for the Gibbs factor of
consecutive Left/Center (Center/Right) blocks at a state $l,c$
$(c,r)$ \cite{KK,KR}. The complexity of the calculation of the block prior
probabilities is $O(Lq^2/ \log q)$ where $L/\log q$ is the number of
blocks.  The decoder complexity per iteration of the MN codes over a
finite field $q$ can be reduced to order
$O(Lqu)$\cite{David-Mackay2,David-Mackay1}, where $u$
stands for the average number of checks per block.  Hence the total
complexity of the DBP decoder is of the order of $O(Lqu+Lq^2/ \log
q)$.

\begin{figure}
\centering
\centerline{\epsfxsize=2.5in \epsffile{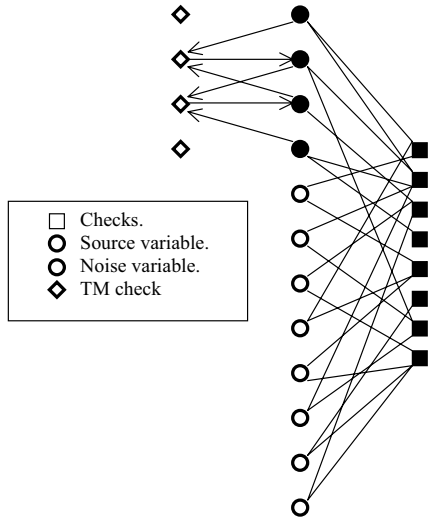}}
\caption{A message passing in the SM joint S/C decoder is represented
by a graph with the following three layers.  The check blocks are
represented by full squares, the full/open circles denote source/noise
block variables and the open diamonds denote the calculators for the
dynamical block posterior probabilities for the source block
variables. Each one of these calculators receives an input message
from its two neighbors (module $L_0$) and sends its output message to
its block.  }
\label{message-passing}
\end{figure}

Another way to represent the dynamical behavior of the SM joint S/C
decoder is in the framework of message passing on a graph. Typically,
the graph is bipartite and consists of variable nodes and check nodes.
A message from variables to checks is a horizontal pass, and a message
from checks to variables is a vertical pass. In the SM joint S/C
decoder there are {\it three} layers, as presented in
Fig. \ref{message-passing}. The first layer represents the checks and
the second layer represents the variables, where each variable and
check stands for a block of $k_0$ bits. The size of the third layer,
denoted as dynamical block posterior probabilities derived from the
Transfer Matrix (TM) method, is equal to the size of the source in
blocks, $L_0=L/k_0$.  Each element in the third layer receives two
arrows, representing the posterior probabilities of the neighboring
blocks, and sends one output arrow to the center block, representing
the current updated dynamical posterior probabilities which are then
used for the vertical pass.


\begin{figure}
\centering
\centerline{\epsfxsize=3.75in \epsffile{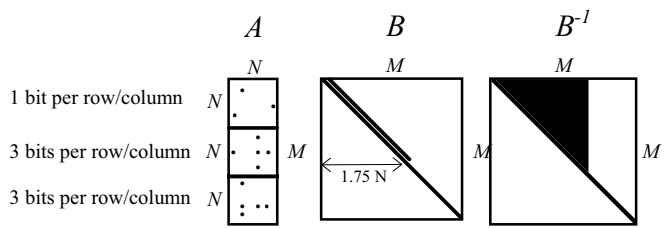}}
\caption{The structure of the matrices $A$ and $B$ for the MN decoder
taken from reference cite{KS}, for rate $1/3$. The black dots (area)
denote the non-zero elements of the matrices $A,~B,~B^{-1}$ }
\label{ks}
\end{figure}

For simplification of the discussion below, in almost all of the
simulation results we concentrate on rate $1/3$ and the construction
of the matrices $A$ and $B$ follow reference \cite{KS} which is
sketched in Fig. \ref{ks}. The advantage of this construction is
that the matrices $A$ and $B$ are very sparse, but the threshold of
the code for large blocks is only $1-3\%$ from the channel
capacity\cite{KS,KS-Gaussian}. Furthermore, since $B$ has a systematic
structure, the complexity of the encoder scales linearly with $L$
although $B^{-1}$ is dense\cite{saad,saad1}. Of course, codes with 
higher thresholds exist (for instance in references
\cite{forney,David-Mackay2}), hence the performance of the joint S/C
algorithm reported below should be interpreted as a lower
bound. (Results for a limited example with rate greater than one,
$R>1$, are briefly discussed in reference \cite{r89})

We conclude this section with the comment that the extension of the SM
joint S/C algorithm in the framework of the MN-Gallager decoder to the
Gallager decoder\cite{Gallager} is in question. In the Gallager
decoder we first solve $L_0(1/R-1)$ equations for the noise variables,
and only in the final step is the message recovered. Since the noise
is not spatially correlated, we do not see a simple way to incorporate
in the Gallager case the side information about the spatial
correlations among the message variables. The equivalence between
these two (MN-Gallager and Gallager) similar decoders is in question.

For illustration, in Fig. 3 we present results for rate $R=1/3$,
$L=10,000$, $q=4$ and $8$ where the decoding is based on the dynamical
block posterior probabilities, eq. \ref{dbp}, and with the following
parameters. For $q=4$ (open circles) $C_1=0.55,~C_2=0.5,~C_{12}=0.4$
($y_1=0.275,~y_2=0.291,~y_{12}=0.422$) and $H_2=0.683$. Shannon's
lower bound, eq. \ref{capacity}, is denoted by the double dotted line,
where for $p_b=0$ the channel noise level is $f_c=0.227$. For $q=8$
(open diamonds) $C_1=0.77,~C_2=0.69,~C_3=0.56,~C_{123}=0.7$
($y_1=0.349,~y_2=0.36,~y_3=0.211,~y_{123}=0.443$) and $H_2=0.453$.
Shannon's lower bound is denoted by the dashed line, where for
$p_b=0$ the channel noise level is $ f_c=0.275$.  Each point was
averaged over at least $1,000$ messages. These results for both
$q=4$ and $8$ indicate that the threshold of the presented decoder
with $L=10,000$ is $\sim 15\%-20\%$ below the channel capacity
for infinite source messages.

\begin{figure}
\centering
\centerline{\epsfxsize=2.5in \epsffile{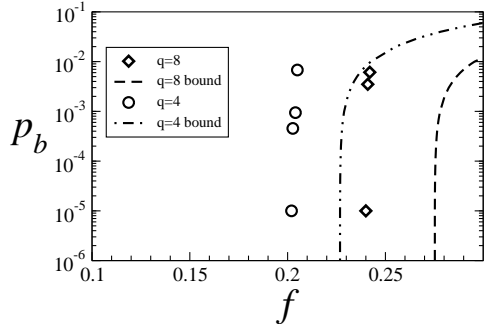}}
\caption{Simulation results for rate $R=1/3$, $L=10,000$, $q=4$ and
$8$ with the following parameters. For $q=4$ (open circles)
$C_1=0.55,~C_2=0.5,~C_{12}=0.4$ ($y_1=0.275,~y_2=0.291,~y_{12}=0.422$)
and $H_2=0.683$. Shannon's lower bound, eq. \ref{capacity}, is
denoted by the double dotted line. For $q=8$ (open diamonds)
$C_1=0.77,~C_2=0.69,~C_3=0.56,C_{123}=0.7$
($y_1=0.349,~y_2=0.36,~y_3=0.211,~y_{123}=0.443$) and $H_2=0.453$.
Shannon's lower bound is denoted by the dashed line.  Each point
was averaged over at least $1,000$ source messages with the desired set
of autocorrelations.  }
\end{figure}

\section{The threshold of the code}

An interesting question is to measure the efficiency of the decoder,
eq. \ref{dbp}, as a function of the maximal correlation length taken
$k_0$, the strength of the correlations, the size of the finite fields
$q$ and to compare the efficiency with the separation schemes.  A
direct answer to the questions raised is to implement exhaustive
simulations on increasing source length, various finite fields $q$,
and sets of autocorrelations, which result in the bit error
probability versus the flip rate $f$. Besides the enormous
computational time required, the conclusions would be controversial
since it is unclear how to compare, for instance, the performance as a
function of $q$; with the same number of transmitted blocks or with
the same number of transmitted bits.

In order to overcome these difficulties, for a given MN-Gallager code
and with DBP decoding over GF(q) and a set of autocorrelations, the
threshold $f_c$ for $L \rightarrow  \infty $ 
is estimated from the
scaling argument of the convergence time, which was previously
observed for $q=2$\cite{KS,KS-Gaussian}.  The median number of
message passing steps, $t_{med}$, necessary for the convergence of the
MN-DBP algorithm is assumed to diverge as the level of noise
approaches $f_c$ from below. More precisely, we found that the scaling
for the divergence of $t_{med}$ is {\it independent of $q$} and is
consistent with
\begin{equation}
t_{med} = {A \over f_c-f}
\label{scaling}
\end{equation}
\noindent where for a given set of autocorrelations and $q$, $A$ is a
constant. Moreover, for a given set of autocorrelations and a finite
field $q$, the extrapolated threshold $f_c$ is independent of $L$, as
demonstrated in Fig. 4.  This observation is essential to determine
the threshold of a code based on the above scaling behavior. Note that
the estimation of $t_{med}$ is a simple computational task in
comparison with the estimation of low bit error probabilities for
large $L$, especially close to the threshold. We also note that the
analysis is based on $t_{med}$ instead of the average amount of
message passing, $t_{av}$,\cite{KS} since we wish to prevent the
dramatic effect of a small fraction of finite samples with slow
convergence or no convergence.\cite{domany,median}

\begin{figure}
\centering
\centerline{\epsfxsize=2.5in \epsffile{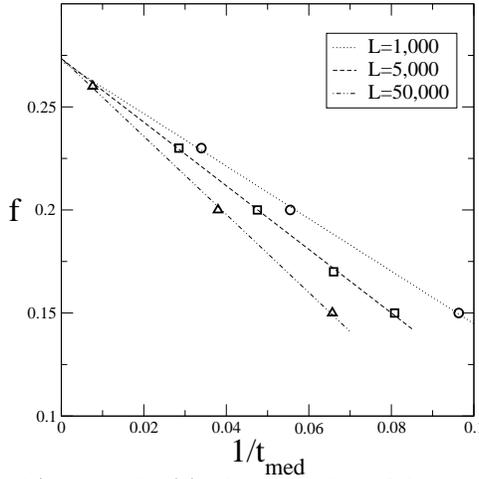}}
\caption{The flip rate $f$ as a function of $1/t_{med}$ for GF(4) with
$C_1=C_2=0.8$ and $L=1,000,~5,000~,50,000$.  The lines are a result of
a linear regression fit. The threshold, $f_c \sim 0.272$, extrapolated
from the scaling behavior eq. \ref{scaling}, is independent of $N$.  }
\end{figure}  

All simulation results presented below are derived for rate $1/3$ and
the construction of the matrices $A$ and $B$ of the MN code are taken
from \cite{KS}. In all examined sets of autocorrelations, $10^3 \le L
\le 5\!\times\!10^4$ and $4 \le q \le 64$, the scaling for the median
convergence time was indeed recovered. For illustration, in Fig. 5, we
present the scaling behavior for the amount of message passing for the
two examined cased presented in Fig. 3. (Note that this decoder can be
extended to rate $R>1$ and results for a limited example are presented
in reference \cite{r89})

\begin{figure}
\centering
\vspace{3.5cm}
\centerline{\epsfxsize=2.5in \epsffile{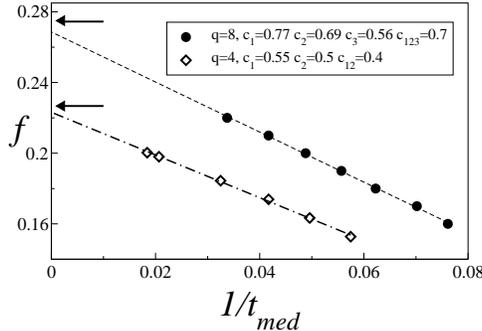}}
\caption{The flip rate $f$ as a function of $1/t_{med}$ for the two
examined cases of Fig. 3. The extrapolated threshold for $q=4,~8$ is
$0.223,~0.265$,  which are about $98\%$ of the Shannon's lower bound 
$0.2267,~0.275$, respectively.
}
\end{figure}

For a given set of autocorrelations, $\{C_{k_1,...,k_m}\}$ where $k_m
\le k_0$, the MN decoder, eq. \ref{dbp}, can be implemented with any
field $q \ge 2^{k_0}$.  In order to optimize the complexity of the
decoder it is clear that one has to work with the minimal allowed
field, $q=2^{k_0}$.  However, when the goal is to optimize the
threshold of the code, the selection of the optimal field, $q$, is in
question. To answer this question we present in Fig. 6 results for
$k_0=2$ ($C_1=C_2=0.86$) and $q=4,~16,~64$. It is clear that the
threshold, $f_c$, increases as a function of $q$ as was previously
found for the case of $i.i.d$ sources.\cite{LDPC-GF(q),KABA} More
precisely, the estimated thresholds for $q=4,~16,~64$ are $\sim
0.293,~0.3,~0.309$, respectively, and the corresponding Ratios
($\equiv f_c/f_{Sh}$) are $0.913,~0.934, 0.962$, where Shannon's lower
bound $f_{Sh}=0.321$.  Note that the extrapolation of $f_c$ for large
$q$ appears asymptotically to be consistent with $f_c(q) \sim 0.316
-0.18/q$.

\begin{figure}
\centering
\centerline{\epsfxsize=2.5in \epsffile{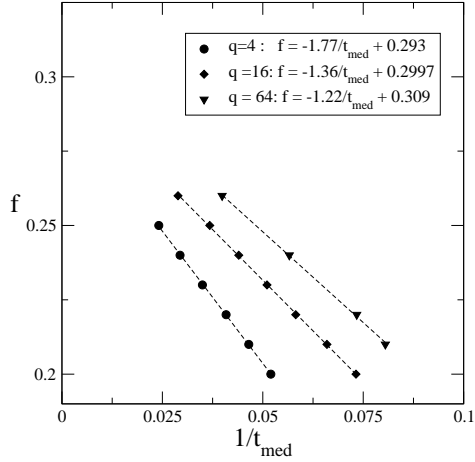}}
\caption{The scaling behavior, $f$ as a function of $1/t_{med}$, for
$C_1=C_2=0.86$ and $q=4,~16,~64$. The lines are a result of a linear
regression fit. The estimated thresholds for $q=4,~16,~64$ are
$0.293,~0.3,~0.309$, and the corresponding $Ratio \equiv f_c/f_{Sh}=
0.913,~0.934, 0.962$, where $f_{Sh}=0.321$.}
\end{figure}

\section{Comparison between joint and separation schemes}

\begin{figure*}
\centering
\centerline{\epsfxsize=5.75in \epsffile{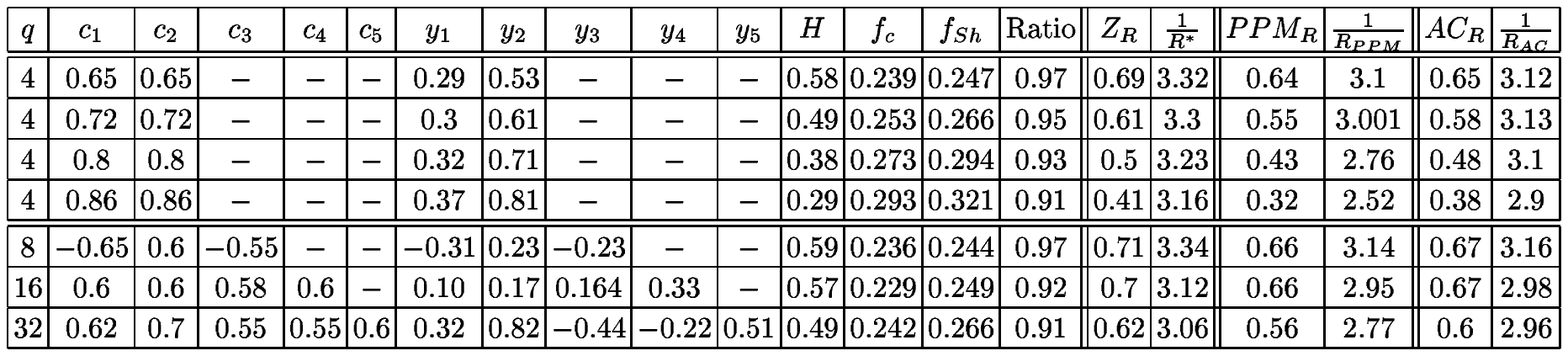}}
\caption{Results for $q=4,~8,~16,~32$ and selected sets of 
two-point autocorrelations $\{C_k\}$
}
\label{t1}
\end{figure*}

Results of simulations for $q=4,~8,~16$ and $32$ and selected sets of
autocorrelations are summarized in Table I (Fig. \ref{t1}) and the
definition of the symbols is: $\{C_k\}$ denotes the imposed values of
two-point autocorrelations as defined in eqs.  \ref{ck} and
\ref{omega}; $\{y_k\}$ are the interaction strengths,
eq. \ref{hamiltonian}; $H$ represents the entropy of sequences with
the given set of autocorrelations, eq. \ref{entropy-ck}; $f_c$ is the
estimated threshold of the MN decoder with the DBP derived from the
scaling behavior of $t_{med}$, eq. \ref{scaling}; $f_{Sh}$ is
Shannon's lower bound, eq. \ref{capacity}; Ratio is the efficiency of
our code, $f_c/f_{Sh}$; $Z_R$ indicates the gzip compression rate
averaged over files of the sizes $10^5-10^6$ bits with the desired set
of autocorrelations. We assume that the compression rate with $L=10^6$
achieves its asymptotic ratio, as was indeed confirmed in the
compression of files with different $L$; $1/R^{\star}$ indicates the
ideal (minimal) ratio between the transmitted message and the source
signal after implementing the following two steps: compression of the
file using gzip and then using an {\it ideal optimal encoder/decoder},
for a given BSC with $f_c$.  A number greater than (less than) $3$ in
this column indicates that the MN joint S/C decoder is more efficient
(less efficient) in comparison to the channel separation method using
the standard gzip compression.  The last four columns of Table I
(Fig. \ref{t1}) are devoted to the comparison of the presented joint
S/C decoder with advanced compression methods. $PPM_R$ and $AC_R$
represent the compression rate of files of the size $10^5-10^6$ bits
with the desired autocorrelations using the Prediction by Partial
Match\cite{PPM} and for the Arithmetic Coder\cite{AC},
respectively. Similarly to the gzip case, $1/R_{PPM}$ and $1/R_{AC}$
denote the optimal (minimal) rate required for the separation process
(first a compression and then an ideal optimal encoder/decoder)
assuming a BSC with $f_c$.

Table I indicates the following main results: (a) For $q=4$ (the upper
part of Table I) a degradation in the performance is observed as the
correlations are enhanced, and as a result the entropy decreases.  The
degradation appears to be significant as the entropy is below $\sim
0.3$ (or for the test case $R=1/3$, $f_c \ge 0.3$).\cite{bias} A
similar degradation was also observed for larger values of $q$ as the
entropy decreases. (b) The efficiency of our joint S/C coding
technique is superior to the alternative standard gzip compression in
the S/C separation technique. For high entropy the gain of the MN
decoder is about $5-10\%$.  This gain disappears as the entropy and
the performance of the presented decoder, eq. \ref{dbp}, are
decreased. (c) In comparison to the standard gzip, the compression
rate is improved by $2-5\%$ using the AC method. A further improvement
of a few percent is achieved by the PPM compression. This latter
improvement appears to be significant in the event of low entropy. (d)
With respect to the performance, the presented joint S/C decoder,
eq. \ref{dbp}, appears to be comparable with the presented separation
methods, but for low entropy it appears that the PPM compression is
superior. However, one should bear in mind a better threshold for the
MN code can be found by optimizing the code \cite{forney}.  (e) With
respect to the computational time of the S/C coding, our limited
experience indicates that the joint S/C decoder is faster than the AC
separation method and the PPM separation method is substantially
slower. Finally, we note that using the side information, the set of
autocorrelations, one can design a special compression procedure which
may overcome the disadvantages of the abovementioned compression
methods \cite{manfred}.

\section{ The role of the spectrum of eigenvalues}

\begin{figure*}
\centering
\centerline{\epsfxsize=5.75in \epsffile{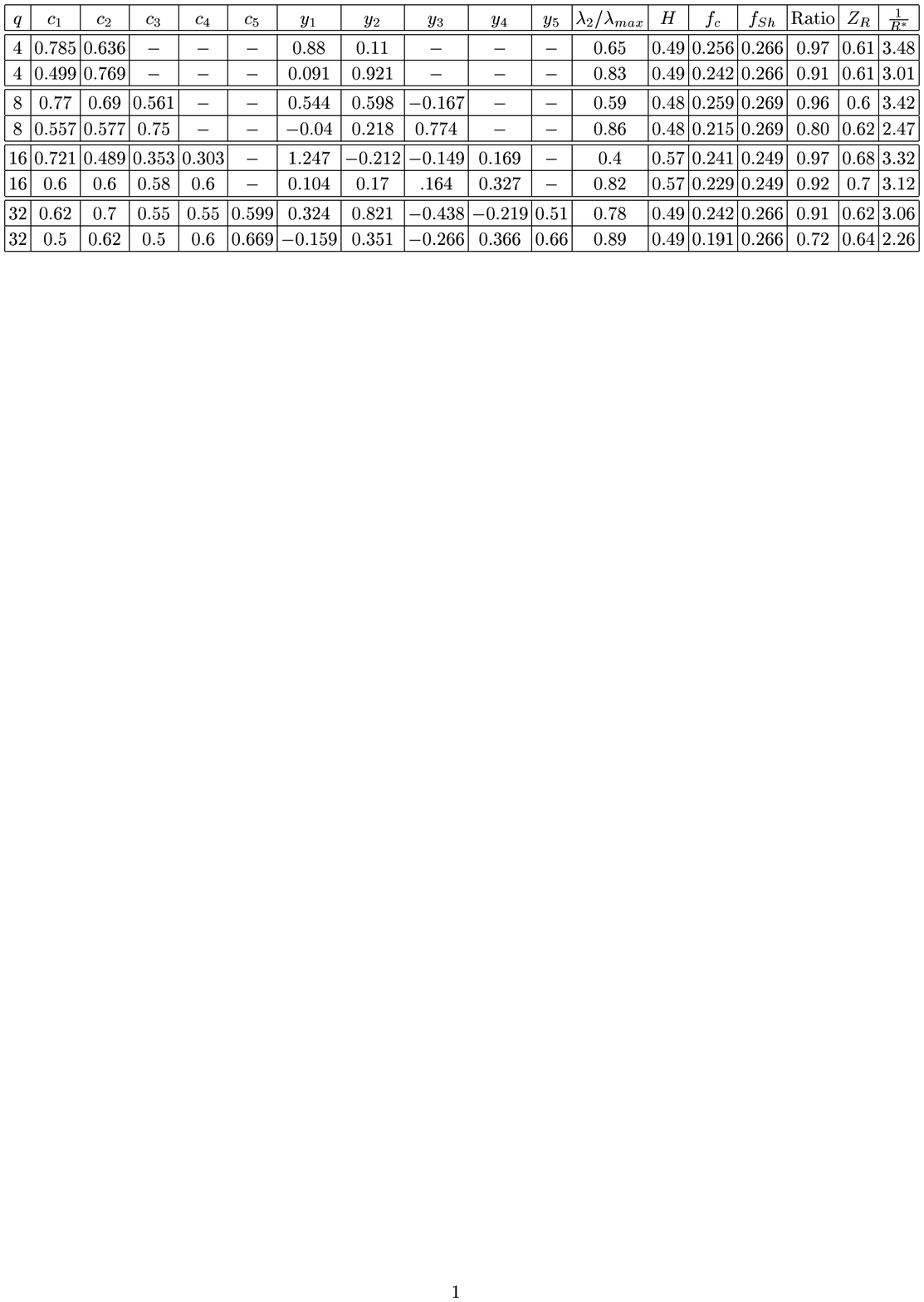}}
\caption{Results for $q=4,~8,~16,~32$ and different sets of two-point
autocorrelations.  For each $q$, two different sets of two-point
autocorrelations characterized by the same entropy and threshold
$f_{Sh}$ are examined.  As $\lambda_2/\lambda_{max}$
increases/decreases, the performance of the joint S/C decoder measured
by the Ratio $f_c/f_{Sh}$ is degradated/enhanced.}
\label{t2}
\end{figure*}

For a given $q$, there are many sets of autocorrelations,
$\{C_{k_1,...,k_m} \}$, in $q$ dimensions obeying the same entropy
(see the discussion in section VI below).  An interesting question is
whether the performance of the presented MN decoder measured by the
Ratio $(\equiv f_c/f_{Sh})$ is a function of the entropy only. Our
numerical simulations indicate that the entropy is not the only
parameter which controls the performance of the algorithm. For the
same entropy and $q$ the Ratio can fluctuate widely among different
sets of correlations.  For illustration, in Table II (Fig. \ref{t2})
results for two sets of autocorrelations with {\it the same entropy}
are summarized for each $q=4,~8,~16$ and $32$. It is clear that as the
Ratio $(\equiv f_c/f_{Sh})$ is much degradated the gzip performance is
superior (the second example with $q=8$ and $32$ in Table II
(Fig. \ref{t2}) where the Ratio is $0.8$ and $0.72$, respectively).
The crucial question is to find the criterion to classify the
performance of the algorithm among all sets of autocorrelations
obeying the same entropy.  Our generic criterion is {\it the decay of
the correlation function over distances beyond two successive blocks}.
However, before examination of this criterion, we return to some
aspects of statistical physics.

The entropy of sequences with a given set of autocorrelations bounded by
a distance $k_0=\log_2(q)$ is determined via the effective Hamiltonian
consisting of $q$ interactions, eq. \ref{hamiltonian}.  As a result
the entropy of these sequences is {\it the same} as the entropy of the
effective Hamiltonian, $H\{y_{k_1,...,k_m} \}$, at the inverse
temperature $\beta=1$, eq. \ref{h2}.  As for the usual scenario of the
transfer matrix method, the leading order of quantities such as 
free energy and entropy are a function of the {\it largest
eigenvalue} of the transfer matrix only. On the other hand the decay
of the correlation function is a function of the whole spectrum of the
$q=2^{k_0}$ eigenvalues (and eigenvectors)\cite{baxter}.
Asymptotically, the decay of the correlation function is determined
from the ratio between the second largest eigenvalue, $\lambda_2$, and
the largest eigenvalue, $\lambda_2/\lambda_{max}$.  From the
statistical mechanical point of view, one may wonder why the first $q$
correlations can be determined using the information of
$\lambda_{max}$ only. The answer is that once the
transfer matrix is defined as a function of $\{y_{k_1,...,k_m} \}$,
eqs. 3-7, {\it all eigenvalues} are determined as well as
$\lambda_{max}$. There is no way to determine $\lambda_{max}$
independently of all other eigenvalues.

In Table II (Fig. \ref{t2}) results of the MN decoder, eq. \ref{dbp}, for
$q=4,~8,~16,~32$ are presented. For each $q$, two different sets of
autocorrelations characterized by the {\it same entropy} and threshold
$f_{Sh}$ are examined.  The practical method we used to generate
different sets of autocorrelations with the same entropy was a simple
Monte Carlo over the space of $\{C_{k_1,...,k_m} \}$\cite{haggai1}.
The additional column in Table II (in comparison with Table I) is the
ratio between $\lambda_2/\lambda_{max}$, which characterizes the decay
of the correlation function over large distances.  It is clear that
for a given entropy as $\lambda_2/\lambda_{max}$ increases/decreases,
the performance of the joint S/C decoder measured by the Ratio
$f_c/f_{Sh}$ is degradated/enhanced, independent of $q$.  The new
criterion to classify the performance of the decoder among all sets of
autocorrelations obeying the same entropy is the decay of the
correlation function.  This criterion is consistent with the tendency
that as the first $k_0$ two-points autocorrelations are
increased/decreased a degradation/enhancement in the performance is
observed (see Table I).  The physical intuition is that as the
correlation length increases, the relaxation time to the equilibrium
macroscopic state increases, and flips on larger scales than nearest
neighbor blocks are required. Finally, we note that in the general
scenario, the first two largest eigenvalues are not sufficient to
approximate the correlation function on short length scales and the
comparison of the efficiency of the decoder should take into account
the entire spectrum of eigenvalues and the eigenvectors \cite{baxter}.

Note that the decay of the correlation function in the intermediate
region of a small number of blocks is a function of all the $2^{k_0}$
eigenvalues.  Hence, in order to enhance the effect of the fast decay
of the correlation function in the case of small
$\lambda_2/\lambda_{max}$, we also try to enforce in our Monte Carlo
search that all other $2^{k_0}-2$ eigenvalues be less than
$A\lambda_{max}$ with the minimal possible constant $A$.  This
additional constraint was easily fulfilled for $q=4$ with $A=0.1$, but
for $q=32$ the minimal $A$ was around $0.5$.

\section{ Possible sets of autocorrelations and the Simplex algorithm}

The entropy of correlated sequences can be calculated from
eq. \ref{h2}.  For the simplest case of sequences obeying only
$C_1$ and $C_2$ the numerical solution of the saddle point equations
indicate that the entropy is non-zero only in the regime
\begin{equation}
-(1+C_2)/2 \le C_1 \le (1+C_2)/2
\label{c1c2}
\end{equation}
\noindent where out of this regime the entropy is zero. The
boundaries, $C_1=|(1+C_2)/2|$, are characterized by the following
phenomena: (a) the entropy falls abruptly to zero at the boundary, and
(b) $y_1$ and $-y_2$ diverge at the boundary (the one-dimensional
Hamiltonian, eq. \ref{hamiltonian} consists of frustrated loops).

These limited results obtained from the numerical solutions of the
saddle point equations suffer from the following limitations: (a) 
finding the boundaries of the regime in the spaces of
$\{C_{k_1,...,k_m}\}$ with a finite entropy is very sensitive to the
numerical precision since on the boundary $\{|y_i|\}$ diverge; (b) it is
unclear whether the available space consists of a connected regime; (c)
the question of whether out of the space with a finite entropy, there are
a finite or infinite number of sequences (for instance $e^{\sqrt{L}}$)
obeying the set of autocorrelations cannot be answered using the
saddle point method; (d) extension of the saddle point solutions
to identify the boundaries of the finite entropy regime to many dimensions 
is a very heavy numerical task.

To overcome these difficulties and to answer the above questions, we
show below how the possible sets of autocorrelations can be identified
using the Simplex algorithm.

For the case of only two constraints $C_1$ and $C_2$, for instance,
let us concentrate on three successive binary variables
$S_i,S_{i+1},S_{i+2}$, where $S_i=\pm 1$.  Since the Hamiltonian,
eq. \ref{hamiltonian}, obeys in this case an inversion symmetry, let us
examine only the $4$ configurations out of $8$ where $S_2=-$,
$(\pm,-,\pm)$. For these $4$ configurations one can assign the
following marginal probabilities, $P(\pm,-,\pm)$, where each
probability stands for the fraction of sequences obeying $C_1$ and
$C_2$ with a given state for these three successive binary
variables. In the SM language we measure the probabilities of these
four states in thermal equilibrium of the micro-canonical ensemble
obeying eq.  \ref{ck}. It is clear that the Hamiltonian,
eq. \ref{hamiltonian}, is translationally invariant,
$P(S_i,S_{i+1},S_{i+2})$ is independent on $i$ after averaging over
all sequences obeying the constraints of eq. \ref{ck}.

For these $4$ marginal probabilities one can write the following
$8$ equations, see eq. \ref{simplex12}.

\noindent For a given $C_1$, these $8$ equations and inequalities can
be solved for the minimum and the maximum available $C_2$ using the
Simplex method. Running over values of $-1 \le C_1 \le 1$, we indeed
recover the result of eq. \ref{c1c2}. However, the {\it Simplex
solution indicates the lack of even finite sequences beyond the regime
with finite entropy}. Hence simple geometrical calculation of
constraint \ref{c1c2} indicates that the fraction of the space
$(C_1,C_2)$ with available sequences is $1/2$.

For the case of three constraints, $C_1, C_2$ and $C_3$, one can
similarly write the following $15$ equalities and inequalities for the
$8$ probabilities of $4$ successive binary variables
$P(\pm,\pm,-,\pm)$, see eq. \ref{simplex123}.

\noindent For a given $C_1$ and $C_2$, these $15$ equations and
inequalities can be solved for the minimum and the maximum available
$C_3$ using the Simplex method. The Simplex solution indicates: (a)
the available solution in the three-dimensional box $(-1:1,-1:1,-1:1)$
for $(C_1,C_2,C_3)$ is a connected region bounded by a few plans whose
detailed equations will be given elsewhere \cite{shahar}; (b) the
fraction of the volume of the box with a positive number of sequences
obeying the three constants is $\sim 0.222$.  Preliminary results
indicate that for $4$ ($C_i,~i=1,2,3,4$) and $5$ ($C_i,~i=1,2,3,4,5$)
constraints the available volume is $\sim 0.085,~0.034$, respectively.

The fraction of possible sets of autocorrelations appears  to decrease as
the number of constraints increases. However, the question of whether the
fraction of available autocorrelations drops exponentially with the
number of constraints as well as its detailed spatial shape is the
subject of our current research \cite{shahar}.

We conclude the discussion in this section with the
following general result \cite{manfred}. The available volume for the
general case of $q$ constraints $\{C_{k_1,...,k_m}\}$ $k_m<\log_2(q)$
is convex. The main idea is that one can verify that the set of
equalities can be written in a matrix representation in the following
form
\begin{equation}
{\bf M} P = C
\label{manfred}
\end{equation}
\noindent where ${\bf M}$ is a matrix with elements $\pm 1$; $P$
represents the marginal probabilities $P(\pm,\pm,....)$ and $C$
represents the desired correlations or a normalization constant
(for instance $C_1/2$, $C_2/2$ and $1/2$, for the case of
eq. \ref{simplex12}). The inequalities force the probabilities into
the range $\lbrack 0:1\rbrack$. Clearly if
$P_1(\pm,\pm,...)$ and $P_2(\pm,\pm,...)$ are two sets of
probabilities obeying eq.  \ref{manfred} then
\begin{equation}
\lambda P_1+(1-\lambda)P_2
\label{convex}
\end{equation}
\noindent is also a solution of the set of the equalities ($0 \le
\lambda \le 1$).  Hence, the available volume is convex.


\begin{eqnarray}
&P&(-,-,+)+P(-,-,-)-P(+,-,+)-P(+,-,-)=C_1/2 \nonumber\\ 
&P&(+,-,-)+P(-,-,-)-P(+,-,+)-P(-,-,+)=C_1/2  \nonumber\\
&P&(-,-,+)+P(-,-,-)+P(+,-,+)+P(+,-,-)=1/2  \nonumber\\
&P&(+,-,+)+P(-,-,-)-P(+,-,-)+P(-,-,+)=C_2/2  \nonumber\\
&0& \le P(\pm,-,\pm) \le 1      
\label{simplex12} 
\end{eqnarray}

\begin{eqnarray}
&P&(+,+,-,+)+P(+,+,-,-)+P(-,-,-,+)+P(-,-,-,-)  \nonumber \\ 
-&P&(+,-,-,+)-P(+,-,-,-)-P(-,+,-,+)- P(-,+,-,-)=C_1/2 \nonumber \\
&P&(+,-,-,+)+P(+,-,-,-)+P(-,-,-,+)+P(-,-,-,-) \nonumber \\
-&P&(+,+,-,+)-P(+,+,-,-)-P(-,+,-,+)-P(-,+,-,-)=C_1/2 \nonumber \\
&P&(+,+,-,-)+P(+,-,-,-)+P(-,+,-,-)+P(-,-,-,-) \nonumber \\
-&P&(+,+,-,+)-P(+,-,-,+)-P(-,+,-,+)-P(-,-,-,+)=C_1/2 \nonumber \\
&P&(-,+,-,+)+P(-,+,-,-)+P(-,-,-,+)+P(-,-,-,-) \nonumber \\
-&P&(+,+,-,+)-P(+,+,-,-)-P(+,-,-,+)-P(+,-,-,-)=C_2/2 \nonumber \\
&P&(+,+,-,+)+P(+,-,-,-)+P(-,+,-,+)+P(-,-,-,-)  \nonumber \\
-&P&(+,+,-,-)-P(+,-,-,+)-P(-,+,-,-)-P(-,-,-,+)=C_2/2 \nonumber \\
&P&(+,+,-,+)+P(+,-,-,+)+P(-,+,-,-)+P(-,-,-,-) \nonumber \\
-&P&(+,+,-,-)-P(+,-,-,-)-P(-,+,-,+)-P(-,-,-,+)=C_3/2 \nonumber \\
&P&(+,+,-,+)+P(+,-,-,-)+P(-,+,-,+)+P(-,-,-,-) \nonumber \\
+&P&(+,+,-,-)+P(+,-,-,+)+P(-,+,-,-)+P(-,-,-,+)=1/2 \nonumber \\
&0& \le P(\pm,\pm,-,\pm) \le 1 
\label{simplex123}
\end{eqnarray}

\section{ Drawbacks of the SM approach}
 
The presented joint S/C decoder based on the SM approach 
suffers from the following drawbacks: 
(a) For each transmitted block one must calculate a $q \times q$
matrix, where each element of this matrix is a function of all $q$
autocorrelations, $\{ C_{k_1,...,k_m}\}$.  Hence, the naive complexity
of the construction of the transfer matrix is $O(q^4)$.  Furthermore,
for each transmitted block the complexity of the calculation of the
leading eigenvalue of the transfer matrix is of $O(q^3)$.  (b) The
required memory is of the order $O(q^2)$, where, for instance, for
$K_0=20$ it results in a 1Mega Bytes.  (c) The solution of the saddle
point equations, eqs. 4-5, requires the calculation many times and
with high precision of the leading eigenvalue of $q \times q$ matrix.
From our experience, the calculation with high precision
of the saddle point equations in $q=2^{k_0}$ dimensions, $\{
y_{k_1,...,k_m} \}$ is a heavy numerical task for $k_0 \ge 4$.  (d)
The extension of the decoder based on the SM approach to include an
array of bits in two or a higher number of dimensions is impossible,
since the trace in eq. \ref{omega} can be done only for very limited
two-dimensional cases\cite{baxter}.

\section{ Joint S/C decoder with advanced threshold}

In order to overcome some of the abovementioned difficulties we
present in this section a decoder with an advanced threshold, where
the decoder gains from fluctuations among different finite source
messages.  For a given sequence of $L$ bits, $\{x_1,~x_2,~...,~x_L\}$,
and $k_m \le k_0$, there are $L_0=L/k_0$ blocks, denoted by
$\{A_1,~A_2,...,~A_{L_0}\}$.  For a given finite field $q=2^{k_0}$ we
denote the number of possible different blocks by
$B_m ~m=1,~2,...,~q$. In the first step of the algorithm, the
probability of occurrence of all three possible successive blocks is
calculated
\begin{equation}
{\hat P}(B_i,B_j,B_k) \equiv {1 \over L_0} \sum_{m=1}^{L_0} \delta_{A_m,B_i}
\delta_{A_{m+1},B_j} \delta_{A_{m+2},B_k}
\label{qqq}
\end{equation}
\noindent where we assume periodic boundary conditions. Note that
although the number of possible triplets of blocks is $2^{3k_0}$, the
complexity of this step for a given source message scales linearly
with $L$.\cite{comment}

The number of non-zero probabilities of occurrence of triplets is
bounded from above by $L_0$ .  However, in a typical scenario of some
enhanced autocorrelations the number of non-zero ${\hat
P}(B_i,B_j,B_k)$ is expected to be $\ll L_0$. Hence in the regime
where $q^3 \gg L_0$ most of the ${\hat P}(B_i,B_j,B_k)$ are equal to
zero, and the tensor, ${\hat P}(B_i,B_j,B_k)$, can be efficiently kept
as a very sparse tensor.  The sparseness of the tensor is expected
even for long sequences, for instance, for $L=10^5$ and $q=128$
($k_0=7$) $128^3 \gg 10^5/7$. In the following we discuss the
importance of this observation.

The decoding of symbols of $k_0$ successive bits is again based on the
standard message passing introduced for the MN decoder over Galois
fields with $q=2^{k_0}$\cite{LDPC-GF(q)} and with the following
modification. The horizontal pass is left unchanged, {\it but a
dynamical set of probabilities assigned for each block is used in the
vertical pass}. The Dynamical Block Probabilities (DBP), $\{P_n^c\}$,
are determined following the current belief regarding the neighboring
blocks in the following way
\begin{equation}
\gamma_{n}^{B_m} = \sum_{i,j=1}^q { {\hat P}(B_i,B_m,B_j) \over \sum_{b=1}^q
{\hat P}(B_i,b,B_j) } q_{m-1}^i q_{m+1}^j
\label{dbp3}
\end{equation}
\begin{equation} 
{\hat P}_{n}^{B_m} = \frac{\gamma_{n}^{B_m}}{\sum _{j=1}^{q}\gamma
_{n}^{j}}
\label{dbp3a}
\end{equation}
\noindent where $q_{m+1}^i/q_{m-1}^j$ stands for the posterior
probabilities of the right/left block in the state $i/j$.

\begin{figure}
\centering
\centerline{\epsfxsize=2.5in \epsffile{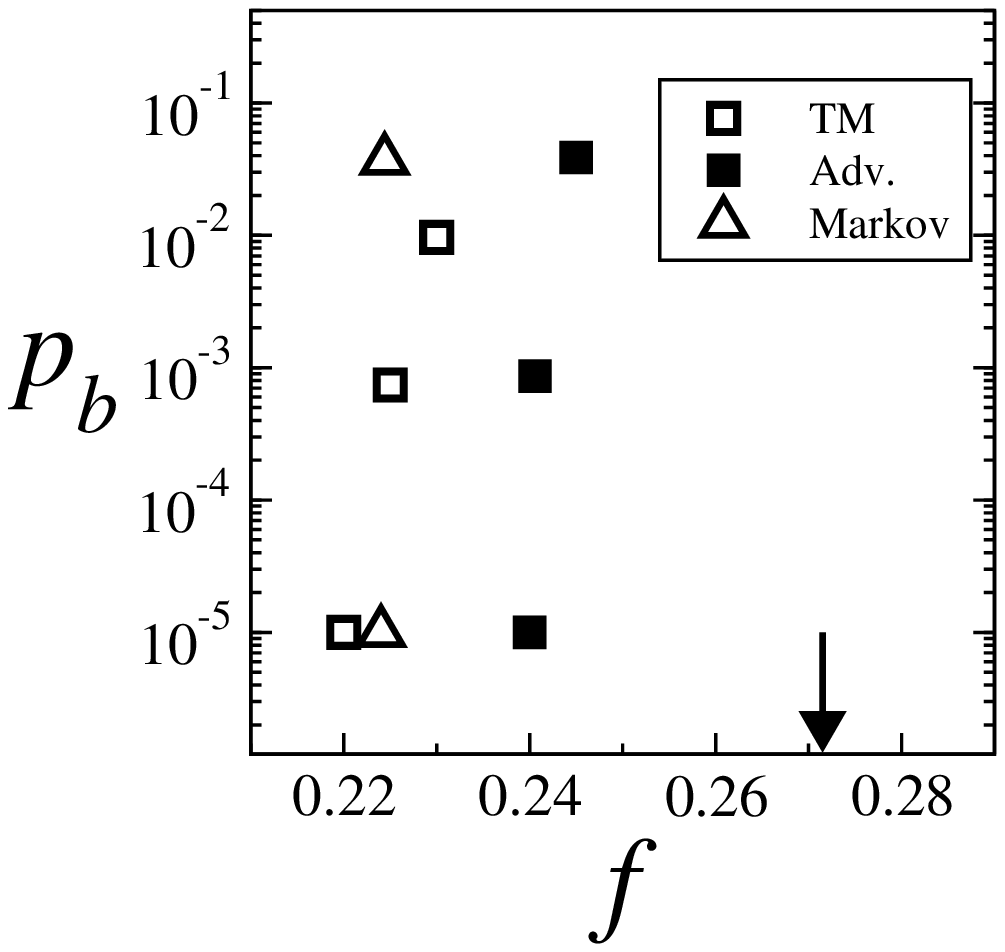}}
\caption{ The bit error rate, $p_b$ versus the channel bit error rate
$f$ for $L=10,000$, $R=1/3$, $q=8$ with
$C_1=C_2=C_3=0.7$. Decoding following the dynamical block
probabilities defined in eq. \ref{dbp} (open squares), decoding
following the advanced joint S/C decoder, eq. \ref{dbp3} (full
squares) and decoding following the Markovian decoder, eq. \ref{pabc}  (open
triangular). Each point is averaged over at least $1,000$ source
messages.  Shannon's lowered bound, $f_c=0.271$, derived from
eq. \ref{capacity} is denoted by an arrow.  }
\vspace{2.5cm}
\label{markov8}
\end{figure}

\begin{figure}
\centering
\vspace{3.5cm}
\centerline{\epsfxsize=2.5in \epsffile{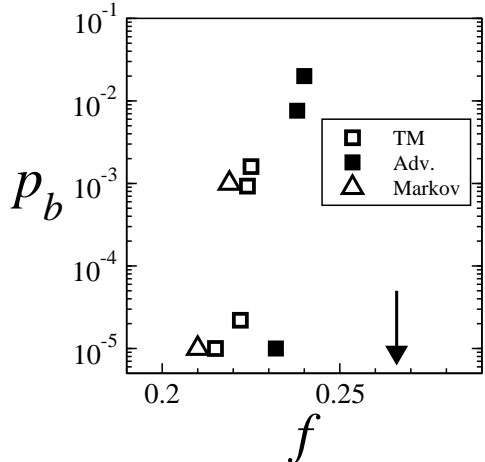}}
\caption{ The bit error rate, $p_b$ versus the channel bit error rate
$f$ for $L=10,000$, $R=1/3$, $q=16$ with
$C_1=0.68,~C_2=0.68,~,C_3=0.6,C_4=0.655$.  Decoding following the
dynamical block probabilities defined in eq. \ref{dbp} (open squares),
decoding following the advanced joint S/C decoder, eq. \ref{dbp3}
(full squares) and decoding following the Markovian decoder,
eq. \ref{pabc} (open triangular). Each point is averaged over at least
$1,000$ source messages.  Shannon's lowered bound, $f_c=0.266$,
derived from eq. \ref{capacity} is denoted by an arrow.  }
\label{markov16}
\end{figure}

We compared the performance of this decoder with the performance of
the previously discussed decoder based on the SM approach,
eq. \ref{dbp}, for different values of $q$ and with rate $1/3$ where
the construction of the matrices $A$ and $B$ again follows
\cite{KS}. Results of the bit error rate, $P_b$, versus the channel
bit error rate, $f$ for $q=8$, and a given set of autocorrelations are
presented in Fig. \ref{markov8}, and for a set of autocorrelations
with $q=16$ in Fig. \ref{markov16} It is clear that the threshold of
the decoder based on eq. \ref{dbp3} is superior to the decoder based
on the SM approach, eq. \ref{dbp}.

Note that for finite $L$ the dynamical block posterior probabilities
defined in eq. \ref{dbp3} (${\hat P}(B_i,B_j,B_k)$) fluctuate among
different samples, where for the decoder based on the SM approach,
eq. \ref{dbp}, these probabilities are sample independent.  This is
one of the sources of the superiority of the presented decoder over
the SM approach (at least for finite $L$).

Note also that the presented decoder, eq. \ref{dbp3}, takes into
account all higher order correlations ($q$ autocorrelations,
$\{C_{k_1,...,k_m}\}$) in a direct measure -- the probability of
occurrence of triplets of blocks, ${\hat P}(B_i,b,B_j) $. There is no
need, as required in the SM approach, eq. \ref{dbp}, to calculate the
form of a $q \times q$ matrix, to diagonalize large transfer matrices
or to seek a saddle point in a large number of dimensions, $q$.

It is clear that the threshold of the advanced joint S/C decoder,
eq. \ref{dbp3}, is superior to the decoder based on the SM approach.
However, from a practical point of view the advanced joint S/C
decoder, eq. \ref{dbp3}, suffers from the following
disadvantages. Firstly, the complexity per message passing scales with
$L_0q^3$ (see eq. \ref{dbp3}), where the complexity of the previously
discussed algorithm is only $L_0q^2$. Secondly, the size of the header
(the transmitted side information, namely, the measured probabilities
of occurrence of triplets of successive blocks) scales also with
$q^3$.  Although the size of the header does not scale with $L$, it is
a critical overhead for a finite $L$.  In the following sections we
show how to overcome these difficulties and to sail towards a
practical algorithm in the large $q$ limit.

\section{Markovian joint S/C decoder}

The calculation of the entropy using the transfer matrix methods
indicates that the ensemble of sequences obeying in the leading order
a given set of autocorrelations can also be derived using a Markovian
process \cite{kinzel}.  More precisely, the elements of the transition
matrix, $\{ P_{ij}\}$ (a transition from state $i$ to $j$), are
related to the transfer matrix elements, $\{ T_{ij} \}$, via the
following normalization
\begin{equation}
P_{ij} = {T_{ij} \over \sum_j T_{ij} }
\label{t-ij}
\end{equation}

Using this analogue, one can now approximate the measured probability
of occurrence of any $q^3$ combinations of three successive blocks,
$(A,B,C)$, using the following formula:
\begin{equation}
{\hat P}(A,B,C) = {{\hat P}(A,B){\hat P}(B,C) \over {\hat P}(B) }
\label{pabc}
\end{equation}
\noindent Hence, the dynamical block probabilities, eq. \ref{dbp}, can
be now calculated with a complexity of $q^2$, and the overall
complexity of the Markovian joint S/C decoder per message passing is
$O(Lq^2/\log(q))$.  Note again that there is no need, as required in
the SM approach, eq. \ref{dbp}, to calculate the form of a $q \times
q$ matrix, to diagonalize large transfer matrices or to seek a saddle
point in a large number of dimensions, $q$

Note that in the limit of infinite $L$, eq. \ref{pabc} is exact in the
leading order of $L$. For a finite $L$, some corrections are expected.
The deviation from a direct measure of the probability of occurrence
of three successive blocks to the estimation of the r.h.s of
eq. \ref{pabc} is expected to be significant only for triplets with
very low probability of occurrence (for instance, if the l.h.s of
eq. \ref{pabc} indicates that a triplet of three successive blocks is
absent in a given sequence where the r.h.s makes one
appearance). However, we do not expect these events with very low
probabilities to dramatically affect the performance of the
algorithm. This expectation was indeed confirmed in our
simulations. Results are exemplified in Figs. \ref{markov8} and
\ref{markov16}, where the performance of the Markovian S/C decoder is
compared with the SM approach, eq. \ref{dbp}.  The difference in the
threshold between these two methods is negligible for the examined
cases.

The complexity of our Markovian S/C decoder was reduced to $O(L_0q^2)$
per message passing. However, there is still a need for the
transmission of the side information consisting of the measured
probabilities of occurrence of all successive pairs of blocks, $\{
{\hat P}(A,B)\}$. Hence the size of the header is of the order of
$O(q^2)$.  For $L \rightarrow \infty$ or more precisely for $L \gg q^2$
the overhead of the transmitted side information is negligible;
however, for a finite $L \le q^2$ it may cancel the benefits of the
Markovian joint S/C decoder.

\begin{figure}
\centering
\centerline{\epsfxsize=2.5in \epsffile{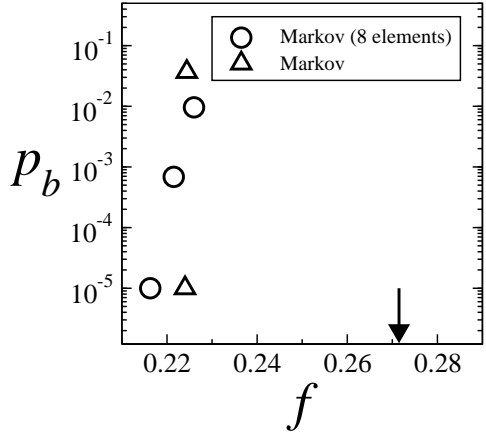}}
\caption{ The bit error rate, $p_b$ versus the channel bit error rate
$f$ for $L=10,000$, $R=1/3$, $q=8$ with $C_1=C_2=C_3=0.7$. Decoding
following the Markovian process, eq. \ref{pabc} (open triangle),
decoding following the Markovian process where only $8$ dominated
elements of the transition matrix, ${\hat P}(A,B)$, are taken as a side
information, and the rest of the elements are set equal to a constant
such that the sum of each row is equal to $1$ (open circle).  
Shannon's lower bound, $f_c=0.271$, is denoted by an arrow.}
\vspace{2.5cm}
\label{q88}
\end{figure}

\begin{figure}
\centering
\vspace{2.5cm}
\centerline{\epsfxsize=2.5in \epsffile{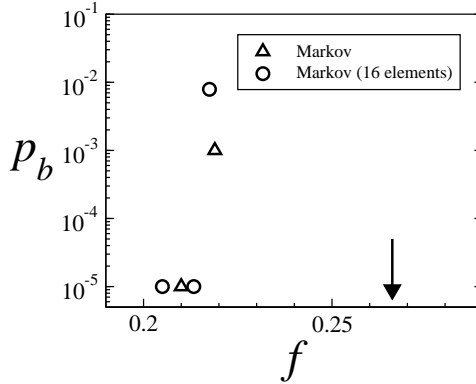}}
\caption{ The bit error rate, $p_b$ versus the channel bit error rate
$f$ for $L=10,000$, $R=1/3$, $q=16$ with
$C_1=0.68,~C_2=0.68,~,C_3=0.6,C_4=0.655$.  Decoding following the
Markovian process, eq. \ref{pabc} (open triangular) and decoding
following the Markovian process where only $16$ dominated elements of
the transition matrix, ${\hat P}(A,B)$, are taken as a side information,
and the rest of the elements are set equal to a constant such that the
sum of each row is equal to $1$ (open circle).  Shannon's lower
bound, $f_c=0.266$, is denoted by an arrow.}
\label{q1616}
\end{figure}

One way to reduce the overhead of the header of the order $O(q^2)$ is
to transmit only the dominated elements of the matrix ${\hat
P}(A,B)$. The remaining elements of the matrix are determined in the
following way. Let us denote the sum of the transmitted dominated
elements in the $i$th row by $M_i$ and their number by $N_i$.
The non-transmitted elements in each row are set equally to
$(1-M_i)/(q-N_i)$. It is clear that as we increase $\{N_i\}$ the
structure of the approximated matrix converges to the true one.  For
sequences with enhanced autocorrelations the structure of the matrix
${\hat P}(A,B)$ was observed to be dominated by a small number of
large elements. The result of simulations for $q=8$ where the number
of transmitted elements, $\sum M_i=8$ (out of $q^2=64$), is presented
in Fig. \ref{q88} and for the case of $q=16$ where $\sum M_i=16$ (out
of $q^2=256$) is presented in Fig \ref{q1616}. The performance seems
to be only slightly affected by this approximation, which dramatically
reduces the required transmitted side information.

An interesting open question is the effect of use of the
sparseness of the tensor ${\hat P}(B_i,B_j,B_k)$ on  the average and
the distribution of the number of required message passing for 
convergence of the decoding process.

\section{Efficient Joint S/C decoder with the lack of side information}

The discussion in the previous sections indicates that the performance
of the presented joint S/C coding is not too far from Shannon's lower
bound and, most probably, using an optimized code (a better
construction for the matrices $A$ and $B$ of the MN code), the channel
capacity can be nearly saturated. However for a finite block length
the main drawback of our algorithm is the overhead of the header which
must be encoded and transmitted reliably. One has to remember that the
size of the header scales with $q^2$ where the precision of each
element is of the order $O(\log L)$. This overhead is especially
intolerable in the limit where
\begin{equation}
{q^2 \log (L) \over L} \sim O(1)
\label{largeblock}
\end{equation}
\noindent Note that this is indeed the situation even for very large
messages, $L=10^6$, and the largest taken autocorrelation length is
only $\log_2q=8$. The l.h.s of eq. \ref{largeblock} with these parameters
is around $1$.

In this section we explain how the abovementioned Markovian joint S/C
can be implemented without the transmission of any side information,
$\{y_{k_1,...k_m}\}$ of eq. \ref{h2}, or ${\hat P}(A,B)$ of
eq. \ref{pabc}.  The main idea can be easily exemplified in the
framework of the Gallager code, where only at the end of the
discussion do we extend it to the MN code.

The first $N$ received bits (the source message) using the Gallager
code with systematic parity-check matrix is the message itself which
is generated by a Markovian process plus the additional channel noise
$f$. Hence, from the receiver point of view the generator of the first
$N$ received bits is a Hidden Markov Process.  The first task of the
receiver is to estimate ${\hat P}(A,B)$ from the knowledge of the
noisy received source message and the channel flip rate $f$.  This
type of {\it parametric estimation} is a common problem in statistics
and can be solved (exactly or approximately) using the EM algorithm or
by one of its variants \cite{EM}. More precisely, for an infinite
source, $L \rightarrow \infty$, the transition matrix eq. \ref{t-ij}
(or the interaction strengths eq. \ref{h2}) can be recovered within a
bounded error with $O(L)$ time complexity. For a finite $L$ sequence
the parameters of the Markovian process can be estimated approximately
with an error of the order $O(q^2/L)$. Hence, the required parameters
for the presented joint S/C decoder based on the dynamical block
posterior probabilities can be estimated from the received noisy
message.  Note that as explained above, the error in the dominated
elements of the transition matrix, ${\hat P}(A,B)$, is the most
important ingredient for the performance of our joint S/C decoder,
hence one may desire an efficient algorithm to estimate especially the
dominated part of the transition matrix.

The critical problem with the above description is that our efficient
decoder can be implemented only by the MN algorithm, where the decoder
{\it simultaneously} estimates the values for the source and noise
bits.  In contrast, in the Gallager decoder, the values for the noise
bits are firstly estimated, and only in the next step are the source
bits recovered. Hence, the dynamical block posterior probabilities
cannot be used (to our current knowledge) in the framework of the
Gallager algorithm.  Since in the MN decoder the source is not
transmitted by itself, the question now is how to estimate the
parameters of the Markovian process which are responsible for
generating of the source message from the received message.
Nevertheless, as explained below, this problem can be solved also
for the MN case in $O(L)$ time complexity.

Let us first explain the solution for the examined MN construction,
Fig. \ref{ks}, then later sketch the general solution.  For the used
MN construction, Fig. \ref{ks}, the first $N$ rows of $A$ are
characterized by one non-zero element per row and column, where the
first $N$ rows of $B$ are characterized by $2$ non-zero elements
(furthermore, each row of $B$ cannot be written as a linear
combination of the other rows). Hence, the first $N$ bits of the
syndrome, eq. \ref{decoding}, are equal to the source with an
effective flip rate equal to $f_{eff}=2f(1-f)$. The EM algorithm with
$f_{eff}$ can now be used to estimate the finite number of parameters
of the Markovian process generating the sequence.

For the general construction of the NM algorithm one adds/subtracts
rows of the concatenated matrix $[A,B]$ and the corresponding received
message bits $Z$ (see eq. \ref{decoding}), such that a situation is
finally reached as follows. The first $N$ rows of $A$ are the identity
matrix, regardless of the construction of the first $N$ rows of $B$, and
with the corresponding $Z_{eff}$.  From the knowledge of the noise
level $f$ and the structure of $i$th row of $B$ one can now calculate
the effective noise level, $f_{i,eff}$, of the $i$th received source
bit. Note that all $N$ effective noise, $\{f_{i,eff}\}$ are  functions
of a unique noise level $f$, and one can again estimate the parameters
of the Markovian process using some variants of the EM algorithm. The
only approximation used in the calculation of $\{f_{i,eff} \}$ {\it
in the general case} is that the new form of the first $N$ rows of $B$
contain loops, hence $\{f_{i,eff} \}$ are correlated. However, these
correlations are assumed to be small as the typical loops are of the
order of $O(\log (L))$.

\section{ Markovian joint S/C coding in higher dimensions}

The decoder based on the SM approach, eq. \ref{dbp}, is limited to a
one-dimensional stream of bits, since the trace in eq. \ref{omega1} can
be done using the transfer matrix method (or any known method) only to
very limited cases of a two-dimensional array of bits \cite{baxter}.
The analytical solution of a two-dimensional Ising system with
arbitrary strength of even nearest neighbor interactions is not
known and in three dimensions no analytical solution is known.  On the 
contrary, the one-dimensional Markovian joint S/C decoder can be
easily extended to a joint S/C coding of a two-dimensional array of bits
or even to an array of bits in higher dimensions.

For illustration, assume that we have a two-dimensional picture to
transmit using a joint S/C mechanism via a noisy channel. A simple way
would be to convert the two-dimensional picture into a one-dimensional
sequence and then to use, for instance, one of the abovementioned
decoders.  However, it is clear that the mapping of the
two-dimensional picture into a one-dimensional sequence is not unique
and of course the natural two-dimensional correlations are destroyed
in this mapping (at least for the realistic case of finite $k_0$). An
alternative way is to generalize the advanced threshold joint S/C
decoder, eq. \ref{dbp3}, to two dimensions, where each block is
updated following its four neighboring blocks.  The generalization of
eq. \ref{pabc} to this case is given by
\begin{eqnarray}
{\hat P}(B_{i,j-1},B_{i-1,j},B_{i,j},B_{i+1,j},B_{i,j+1}) \equiv 
\nonumber \\
{1 \over L_0^2} \sum_{i,j}^{L_0} \prod_{k+m=-1,0,1}
\delta_{A_{i+k,j+m},B_{i+k,j+m}}
\label{pabc-2d}
\end{eqnarray}
\noindent where $L_0^2=(L/k_0)^2$ is the number of blocks of the
two-dimensional array of bits, and again periodic boundary conditions
are assumed.  Similarly to eq. \ref{pabc}, the dynamical posterior
probabilities now take the following form
\begin{eqnarray}
\gamma_{n}^{B_{k,m}} = \sum_{i,j,s,t=1}^q {
{\hat P}(B_{i},B_{j},B_{k,m},B_{s},B_{t}) \over \sum_{b=1}^q
{\hat P}(B_i,b,B_j) }\times  \nonumber \\ q_{k-1,m}^{i} q_{k,m-1}^j
q_{k+1,m}^{s} q_{k,m+1}^t
\label{gamma-2d}
\end{eqnarray}
\noindent It is clear that the generalization of this decoder to
a higher dimension is straightforward and in the naive decoding the
complexity of the decoder scales as $L_0^dq^{2d-1}$.  Nevertheless,
similarly to the Markovian joint S/C decoder also in the higher
dimensional case the complexity can be reduced and, for instance, in two
dimensions
\begin{eqnarray}
&{\hat P}(B_{i,j-1},B_{i-1,j},B_{i,j},B_{i,j+1},B_{i+1,j})= \nonumber \\
&{ {\hat
P}(B_{i,j-1},B_{i,j}) {\hat P}(B_{i-1,j},B_{i,j}) {\hat
P}(B_{i,j+1},B_{i,j}) {\hat P}(B_{i+1,j-},B_{i,j}) \over {\hat
P}(B_{i,j})^3 }
\label{pabc-d}
\end{eqnarray}
\noindent and similarly in higher dimensions. Hence, the complexity of
a message passing in $d$ dimensions is reduced to $(L_0)^dq^2$, or
alternatively the complexity per block is of the order of $O(q^2)$.

Besides the above simplification, it is important to note that for
finite $L$ the tensor of the probabilities of occurrence of nearest
blocks, for instance eq. \ref{pabc-d}, for the two-dimensional case,
is expected to be very sparse. Hence, the decoder can be accelerated
as was discussed for the one-dimensional case.

From an analytical point of view, we do not have an effective way to
generate an ensemble of arrays with a given set of autocorrelations in
two or higher dimensions, since we do not know how to derive the
effective interactions, eq. \ref{hamiltonian}. From a practical point
of view, for a given two-dimensional picture and $k_0$, we can
measure the correlations, eq. \ref{pabc-d}, and then apply the
Markovian decoder. However, there is no reference point to compare the
efficiency of the decoder, since we do not have an effective way to
calculate the entropy, $H_2$, and then Shannon's lower bound,
eq. \ref{capacity}, for a given set of correlations in more than
one dimension. Practically, 
the performance of the Markovian decoder in higher dimensions can be
compared to other known efficient lossless compression methods for two
and higher dimensions.  This important comparison certainly warrants 
further research.

\section{Concluding remarks}

The only remaining major drawback of the presented Markovian joint S/C
coding is that the complexity of the decoder scales in the leading
order for large $q$ per message passing as $O(Lq^2/\log_2(q))$. 

We note that asymptotically the complexity of the Markovian joint
S/C decoder per message passing might be reduced to $O(Lq\log(q))$.  The
main idea can be exemplified in the framework of the original SM
scheme, eq. \ref{dbp}. The complexity of the calculation of each
$\gamma_n^c$ is of the order of $q$, and it is required to calculate
such $q$ different elements. Each summation in eq. \ref{dbp} consists
of the following two types of terms. The first one is the static
terms, $S_{L}(l,c) (S_R(c,r))$, which are the Boltzmann factors, or a
row of the transition matrix of the Markov process. The second
type is the dynamical posterior probabilities for the neighboring
blocks, $q_L^l,~q_R^r$.  The static terms can be ordered in a decreasing
rank order only once, in the initial stage of the decoder, and the
first $O(\log(q))$ largest dynamical posterior probabilities can be
found at the cost of $O((q\log(q)))$ per block.  Next we run the usual
decoder, eq. \ref{dbp}, in one of the following two options: (a) the
summations in eq. \ref{dbp} are done only on the {\it current} leading
$O(\log(q))$ of the dynamical block posterior probabilities, or
alternatively (b) the summations are done as was proposed in (a) with
the additional $O(\log(q))$ leading terms of the static terms, or any
combination of (a) and (b).

The idea behind the above procedure is similar to results of section
IX, where only a limited degradation in the performance of the
Markovian decoder was observed where the transition matrix was
approximated by the knowledge of only a small number dominated
terms. Similarly in the presented approximation, we expect that most
of the bits are correctly ordered by the dominated part of the
static Boltzmann weights and by the dominated part of the dynamical
block posterior probabilities.  In the final stage of the decoder,
rare events of a pair of blocks, small Boltzmann factors, will be
correctly ordered by the dominated true posterior block probabilities
for one of its two neighbors.

The question which remains to be answered is the origin of the
suggested scaling of the order of $O(\log(q))$ dominated taken terms
in the summations of eq. \ref{dbp}. The explanation is based on the
characteristic features of random graphs \cite{erdos,kanter1}. In the
full operation of the Markovian process one assigns for each pair of
nearest blocks a dynamical transition matrix of size $q \times q$,
which resembles a fully connected graph consisting of $q$ nodes. The
purpose of our approximation is to replace the fully connected graph
with the diluted one, {\it but the graph has still to be connected};
the maximal component of the graph must be $q$. The lack of finite
components is a necessary condition, since in such an event the
enhancement of the true block posterior probability may be dynamically
forbidden, since there are isolated nodes (states).  From the random
graph theory it is know that the maximal component is equal to $q$
where the average connectivity (the average number of non-zero
transitions per row) is of the order of
$O(\log(q))$\cite{erdos,kanter1}. This prediction has still to be confirmed
in large scale simulations, large $L$ and $q$.

Finally, note that for large $q$ the transition matrix, ${\hat
P}(A,B)$, is a very sparse matrix in the limit $q^2 \gg L$. This limit
is achieved even for very large source messages and short-range
correlation length, for instance, $k_0=12, ~q=2^{k_0}=2048$ and
$L=10^5$. Furthermore, in the limit where the number of possible
different blocks $q=2^{k_0} \gg L$, a large fraction of $\gamma_n^c$,
eq. \ref{dbp}, can be taken as zero probabilities. Hence, the
complexity of the decoder can be simplified further by these two effects.

\section*{Acknowledgment}

I.K thanks David Forney, Wolfgang Kinzel, Manfred Opper, Shlomo
Shamai, Rudiger Urbanke and Shun-ichi Amari for many helpful
dicussions and comments.


\begin{thebibliography}{1}
\bibitem{median} In practice we define $t_{med}$ to be the average
convergence time of all samples with $t \le$ the median time.
\bibitem{r89} For rate $9/8$, for instance, the chosen construction
for the matrices $A$ and $B$ are as follows. For $A$, the fraction of
rows (from the first row of $A$) $(1/16,1/4,9/16,1/16,1/16)$ are
characterized by $1,2,3,5,9$ non-zero elements per row, respectively.
The structure of $B$ is the same as illustrated in Fig. \ref{ks}, but
$1.75$ is replaced with $7/9$. We ran simulations for this
construction with $C_1=C_2=0.7$ and the corresponding entropy is
$H_2=0.513$ and $L=9,000$.  The extrapolation of $t_{med}$ indicates
that the threshold of this code for large $L$ is $f_c \sim 0.057$. In
the separation scheme using {\it optimal compression and error
correction schemes} and with $f_c=0.057 ~(R_{i.i.d}=0.618)$, one can
find that the overall inverse rate of the communication channel is
$1/R=0.513/0.618 \sim 0.83$, which is only about $6\%$ below our joint
S/C inverse rate $1/R=8/9 \sim 0.89$.  One must remember that our MN
construction can be further optimized, and the critical channel noise
is expected to be enhanced, $f_c > 0.057$.
\bibitem{bias} A similar degradation in the performance was observed
for $q=2$ and biased binary messages (each source bit is equal $0/1$ and is
chosen with probability $p/1-p$).  As $|p-0.5|$ increases the
entropy decreases and a degradation in the performance of the MN
algorithm was observed.
\bibitem{comment} In principle one can generalize the tensor to
include nearest and next-nearest blocks.
\end{thebibliography}
\end{document}